# Applications of the MRI-model in Sputter Depth Profiling

Siegfried Hofmann

National Research Institute for Metals
1-2-1 Sengen, Tsukuba, Ibaraki 305 Japan
e-mail: siegho@nrim.go.jp

The physical principles of the MRI model (abbreviation for mixing-roughness-information depth) for the theoretical calculation of measured concentration-depth profiles obtained in sputter depth profiling with AES, XPS, SIMS etc. are outlined. The main features of the calculation and the respective visual Basic program are explained. Examples demonstrate the usefulness of the current approach for the reconstruction of the original in-depth distribution of composition.

1. **Introduction**

The aim of any profiling experiment is to get a sputter depth profile (intensity as a function of sputtering time) which resembles as closely as possible the original elemental distribution with depth. However, a somehow distorted image of the latter is generally obtained even when the experimental setup is carefully optimized. Therefore, quantification is necessary which requires at least three steps: (I) Quantification of the depth scale (sputtering rate), (II) Quantification of the concentration scale (relative sensitivity factors etc.) and (III) Assessment of the depth resolution [1,2]. Whereas the depth resolution in the usual definition of $\Delta z$ (taken as the depth difference between 16% and 84% of the plateau value [3]) is a quantitative measure of the depth range within which we only have an average concentration value and cannot tell anything about details of concentration distribution directly from the experimental data, deconvolution or reconstruction of the in-depth distribution is the final step in depth profiling [4]. To perform this task, we have to know the depth resolution function either by experiment or by theoretical prediction [1-6]. One of the first model predictions of the latter was the so-called SLS (Sequential Layer Sputtering) model based on the increase in surface roughness with sputtered depth according to a Poisson function [1]. Although it has been questioned for overestimating roughness and at the same time neglecting atomic mixing, the underlying mathematics are practically the same at the beginning of sputtering near the surface, and the SLS model was found very useful for profile reconstruction for the first 20 monolayers as e.g. for oxidation or passive layers in corrosion [7].

For thicker metallic layers, roughness is generally the dominating contribution to $\Delta z$ [8]. Particularly since the introduction of sample rotation during profiling [9], sputtering induced roughening due to crystalline orientation and/or surface reaction dependent local variation of the sputtering rate can now be avoided and high resolution depth profiling with a depth resolution $\Delta z$ below 5 nm is generally achieved when low energy ions ($\leq$1keV Ar+) are used [1]. Below about $\Delta z=5$ nm, atomic mixing and information depth are becoming of increasing importance. Ultra-high depth resolution with $\Delta z< 1.5$ nm (a few atomic monolayers) is expected for the case of low ion energy and/or high incidence angle, and by using low energy energy Auger electrons (<100 eV) in AES depth profiling. Only if these experimental parameters are pushed to the limit, ultra-high depth resolution is attained [10], for which so far only a few examples exist [11,12]. The ultimate limit seems to be about 3-4 monolayers ($\Delta z=0.7-1.0$ nm, [13]).

Features smaller than Δz, e.g. of the order of an atomic monolayer, can only be resolved by deconvolution or profile reconstruction procedures with an appropriate depth resolution function. Accurate experimental determination and theoretical modeling of the depth resolution function are necessary for a correct transformation of the measured sputter profile into the original depth distribution of composition. Such a high accuracy was recently achieved by introduction of the so called MRI-model, for atomic mixing (M), surface roughness (R) and information depth (I) for the description of the depth resolution function.

The MRI model was developed by the end of the eighties in the Applied Surface and Interface analysis group at the Max-Planck-Institut fuer Metallforschung in Stuttgart [14,15] and is described in detail in ref. [16]. A program in Q-Basic was available since that time and was distributed to some researchers on request. Recently, a more comfortable and user-friendly version in Visual Basic was developed by Schubert and Hofmann [17] at the National Research Institute for Metals in Tsukuba. After careful debugging and further improvement it is planned to give it to the Surface Analysis Society for practical use.

In this paper, a few recent applications of the MRI-model to AES and SIMS depth profiles will be discussed in order to demonstrate the usefulness of the approach as well as its capabilities and limitations.

2. **Outline of the MRI- model**

According to the well known convolution integral which describes the folding of the in depth distribution of composition, X(z'), with the normalized depth resolution function g(z-z'), the resulting normalized intensity I(z)/I(0) is represented by:

@ @ @ @ @ @

$$I(z)/I(0) = \int_{-\infty}^{+\infty} X(z') \cdot g(z-z') dz' \quad @(1)$$

For X(z') being a delta function, the measured I(z)/I(0) is identical with the depth resolution function. The latter can also be obtained by differentiation of the measured profile at a sharp interface, as seen from eq. (1), when d((I(z)/I(0))/dz = g(z-z') [2,5,6].

Deconvolution of a measured profile after eq. (1) means its solution for X(z') with the knowledge of I(z)/I(0) and g(z-z'). such a "reverse" problem often faces practical difficulties because of isufficient precision, low number of data points and of insufficient signal-to- noise level. Therefore it has become customary to solve the problem by "forward calculation" using eq. (1) directly, i.e assuming a suitable X(z') and comparing the calculated I(z)/I(0) with the measured profile. Then making input changes until an optimum fit by trial and error is obtained [4].

The MRI- model is capable of giving a mathematical description of the depth resolution function g(z-z'), based on the three fundamental parameters atomic mixing, surface roughness and information depth, i.e. g(z-z') = {g(w);g( →);g( )}. Atomic mixing is described by an exponential function with a characteristic mixing zone length, w, the information depth by another exponential function with a characteristic length , and the roughness by a Gaussian term with standard deviation σ (corresponding to rms roughness). These functions are employed sequentially to the (assumed) depth distribution of an element, given by thin layers each with (different) concentrations. For example, each monoatomic layer at a location $z_0$ gives a normalized contribution at a sputtered depth z which is described by

atomic mixing:
$$g(w) = \exp[-(z - z_0 + w)/w] \quad (2)$$

information depth:
$$g(\lambda) = \exp(-(z - z_o)/\lambda) \quad @ @(3)$$

surface roughness:
$$g(\sigma) = \frac{1}{\sqrt{2\pi}\sigma} \cdot \exp\left[\frac{-(z-z_0)^2}{2\sigma^2}\right] \quad @ (4)$$

Eqns. (2,3,4) can be applied by summing up all the contributions for each depth z after eq. (1), thus representing the calculated depth profile which can be compared to the measured one. This is shown in the examples below (see Fig. 1)
.

## 3. Application Examples
### 3.1 AES depth profiling

Besides the authors' group, at least three other groups have reported good success when applying the MRI model to quantify their data [5,12,17]. Excellent results were obtained with GaAs/AlAs multilayer structures [16, 18]. An example for a structure consisting of 40 ML (11.2 nm) GaAs, 1 ML AlAs, 40 ML GaAs and 36 ML AlAs on GaAs is presented in Fig.1, where the experimental and the calculated AES profiles of the Al (1396 eV) peak-to peak-height are plotted and shown on the visual basic template for the current version MRI 2.1 together with the reconstructed in depth distribution of Al (below these curves in block picture). Most of the program-surface was designed to be self-explaining. Input of a layer-structure can be done either via a data file or manually. In each line of the data file the depth and concentration at each concentration change within the structure should be written.

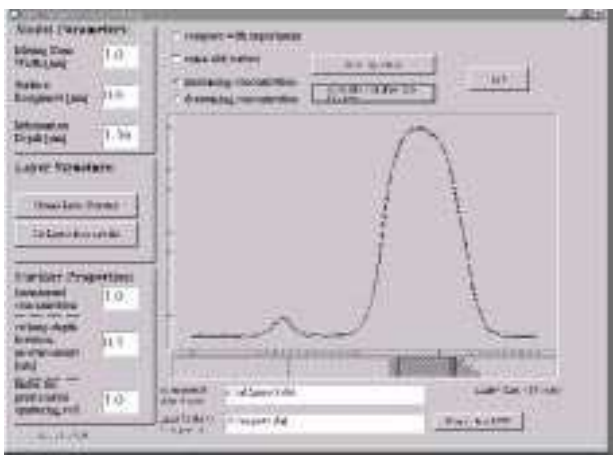

**Fig. 1**: Visual Basic surface of the MRI model (Ver. 2.1) with measured data, fitted profile and reconstructed depth distribution of the example described in the text.

For comparison with experimental data, the data-file should contain in each line depth z and measured intensity I (for example AES-peak-to-peak-height). The plot of the data curve is normalized to fit into the drawing window, the profile-height can be reduced by setting the maximum concentration to a value smaller than 1.The etching depth between the measurements is given by the sputtering rate. Exported will be the data-set from the last calculation in the form $z(i)$, $c(i)$. The factor for preferential sputtering is at present only for the value $z(i)$ which will no longer represent the true sputtered depth, but a value proportional to the sputtering time. It is currently under work for improvement.

In the case of Fig.1, depth profiling was done with 500 eV Ar+ ions at about 68 degree incidence angle. A good fit of experimental and calculated results with the depth resolution function derived from the monolayer profile and fitted with the MRI-parameter shown in the upper left of Fig.1: mixing zone thickness (w = 1.0), roughness ( $-$=0.6 nm) and information depth ( =1.16 nm) can only be obtained if the first interface (GaAs/AlAs) is slightly changed and the second interface (AlAs) is considerably changed in depth distribution of composition [18]. This demonstrates the sensitivity of the model.

### 3.2 SIMS Depth Profiling

The main difference between SIMS and AES is for preferential sputtering, because SIMS detects the sputtered matter whereas AES (and XPS) detects the altered surface layer. For negligible preferential sputtering, there is no principal difference except in the analysis signal. Whereas the roughness ( $-$) and mixing (w) terms are therefore identical in AES and SIMS, the information depth ( ) is principally different because it depends on the analyzed signal. In AES it can be selected with respect to different kinetic energy and emission angle of the Auger electrons. The information depth in SIMS is given by the secondary ion emission depth, which is of the order of an atomic monolayer and shows almost no influence of the primary ion energy. Therefore we set =0.4 (also sucessfully used for other SIMS profiles, see e.g. [13]) for the MRI model application to the Si$^+$ ion SIMS depth profile of an SiO$_2$/Ta$_2$O$_5$ multilayer structure shown in Fig. 2. The SIMS data were kindly provided by Dr. Moon [19].

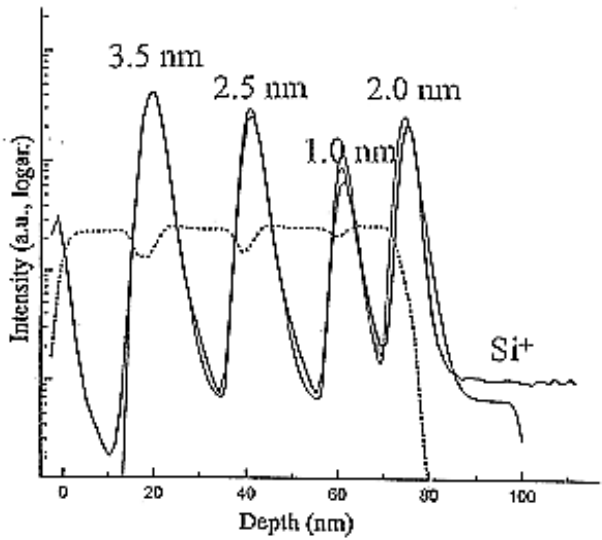

**Fig. 2**: Si$^+$ depth profile (SIMS) of SiO$_2$ layers in Ta$_2$O$_5$, measured [19] and MRI-fitted data (see text).

The mixing length w is of the order of the primary ion range and is determined by the characteristic decay length of the downward slope of the Si$^+$ profile, which gives w= 1.6 nm. The roughness parameter is varied until an optimum fit is obtained for $-$= 1.0 nm. However, to achieve optimum fit, the thicknesses of the 2$^{nd}$ and 3$^{rd}$ SiO$_2$ layers has to be changed from the nominal value given in the Figure to 2.2 nm (from 2.5 nm) and to 0.6 nm (from 1.0 nm), and furthermore a small, constant background of 1.1 ~10$^{-3}$ of the Si$^+$ intensity for thick SiO$_2$ layers is necessary to yield the depth and approximate curvature of the profile between the layers (this can be explained by crater edge, redeposition and resputtering effects). The additional, less well fitting line in the 3$^{rd}$ layer is obtained when changing the respective layer thickness from 0.6 to 0.8 nm, thus demonstrating the high accuracy of less than a monolayer. No good fit is expected for the 4$^{th}$ layer because this layer is on a different matrix (Si substrate).

## 4. Conclusion

Numerous examples of succesful applications of the MRI model in SIMS and AES depth profiling have shown its capability to provide the reconstruction of the original in-depth distribution of composition. Of course, the ultimate limit is set by the signal to noise figure of the experimental data. However, more important limitations at present are the simplified model assumptions and the degree of realization in any specific experiment. For example, a mutual profile shift in GaAs/AlAs multilayer depth profiling with AES high and low energy Al peaks was found to be about one monolayer larger than predicted by the MRI model [16], but could be correctly simulated by Monte Carlo simulations which take into account the inhomogeneous distribution within the mixing zone [20]. In favorable cases, i.e. when no nonlinear effects with concentration and sputtering rate occur, monolayer accuracy can be obtained. Whereas SIMS, due to its high sensitivity, is well suited for low concentration delta layers and dopant profiles, AES is much less sensitive but shows comparatively little matrix effects and is therefore better suited for interfacial profiling. Nonlinear effects can de detected by profiling reference samples which contain a delta layer and an interface of the same element in the same sample as shown in Fig.1. In principle, nonlinear effects like preferential sputtering can be taken into account. However, compositional distriibution within the mixing zone determines the accuracy of profile reconstruction. Therefore, basic limitations are encountered when preferential sputtering occurs in combination with radiation enhanced diffusion and segregation [21].